\documentstyle[fleqn,run2col,epsf]{article}
\epsfverbosetrue
\catcode`@=11
\def\citer{\@ifnextchar
[{\@tempswatrue\@citexr}{\@tempswafalse\@citexr[]}}

\def\@citexr[#1]#2{\if@filesw\immediate\write\@auxout{\string\citation{#2}}\fi
  \def\@citea{}\@cite{\@for\@citeb:=#2\do
    {\@citea\def\@citea{--\penalty\@m}\@ifundefined
       {b@\@citeb}{{\bf ?}\@warning
       {Citation `\@citeb' on page \thepage \space undefined}}%
\hbox{\csname b@\@citeb\endcsname}}}{#1}}
\catcode`@=12
\newcommand{\beq}{\begin{equation}}
\newcommand{\eeq}{\end{equation}}
\newcommand{\beqa}{\begin{eqnarray}}
\newcommand{\eeqa}{\end{eqnarray}}

\def\simgt{\rlap{\lower 3.5 pt \hbox{$\mathchar \sim$}} \raise 1pt \hbox
{$>$}}
\def\simlt{\rlap{\lower 3.5 pt \hbox{$\mathchar \sim$}} \raise 1pt \hbox
{$<$}}

\newcommand{\np}[3]{{Nucl.\ Phys.} {\bf #1} (19#2)~#3}
\newcommand{\pl}[3]{{Phys.\ Lett.} {\bf #1} (19#2) #3}
\newcommand{\pr}[3]{{Phys.\ Rev.} {\bf #1} (19#2) #3}

\def \MSbar {\vbox{\hrule\kern 1pt\hbox{\rm MS}}}

\def\fig#1{Fig.~\ref{fig:#1}}
\def\tab#1{Table.~\ref{tab:#1}}
\def\junk#1{}

\def\tabgkl{
\begin{table}[hbtp]
\begin{tabular}{||l||c|c|c|c||} \hline \hline
 PDF Set           &Mass (GeV)    & LO    & $WQ\bar{Q} $  &  NLO \\   \hline \hline
  CTEQ1M        &$m_c$=1.7      & 96    &20             &161\\  \hline
  MRSD0'        &$m_c$=1.7      & 81    &20             &138\\  \hline
  CTEQ3M        &$m_c$=1.7      & 83    &20             &141\\  \hline
  CTEQ3M        & $m_b$=5.0     & 0.17   &9.09           &9.33 \\  \hline \hline
\end{tabular}
\vskip -0pt
\caption{
The $W$ + charm-tagged one-jet inclusive cross section in $pb$ for
LO, $W+Q\bar{Q}$, and NLO (including the $W+Q\bar{Q}$ contribution)
using  different sets of
parton distribution functions. Table is taken from 
Ref.~\protect\cite{gkl}.
\label{tab:gkl} 
}
\end{table}
}
\def\figband{
\begin{figure}[ht] 
\begin{center}
\leavevmode
 \epsfxsize=0.85\hsize \epsfbox{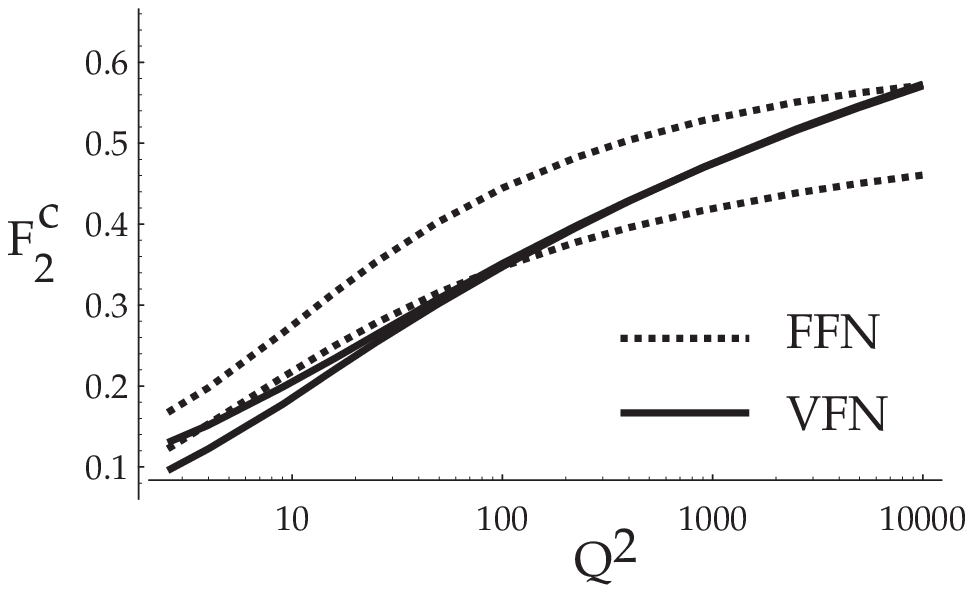} 
\vskip -20pt
\caption{
 $F_2^c$  for $x=0.01$ as a function of $Q^2$  in GeV 
 for two choices of $\mu$ 
 as obtained within the ${\cal O}(\alpha_s^1)$
 FFN and (ACOT) VFN schemes. 
 For details, see Ref.~\protect{\cite{schmidt}}.
\label{fig:band} 
}
\vskip -20pt
\end{center}
\end{figure}
}
\def\figsacot{
\begin{figure}[ht] 
\vbox{
 \hbox{
 \epsfxsize=0.85\hsize \epsfbox{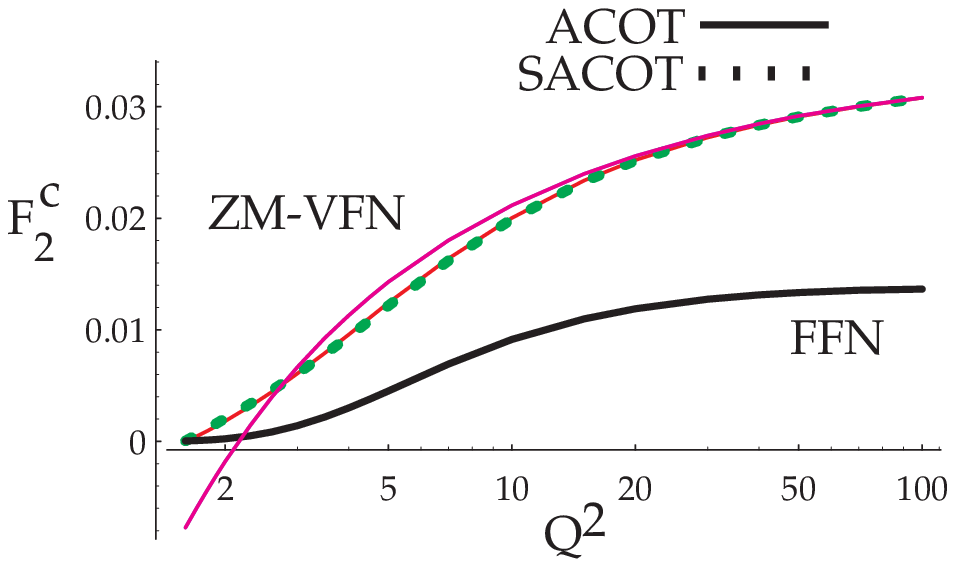} 
 }
 \hbox{
 \epsfxsize=0.85\hsize \epsfbox{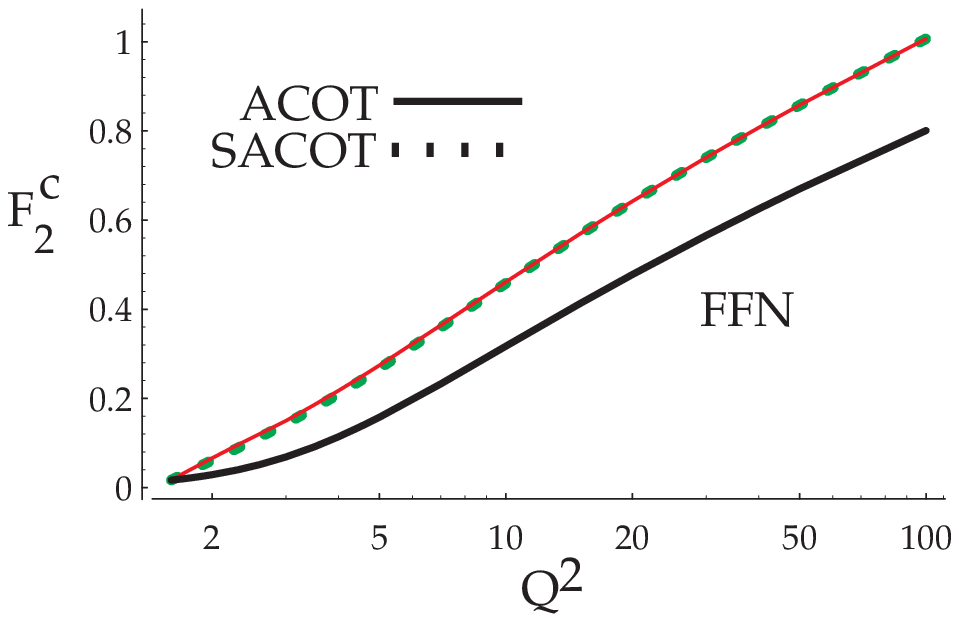}
 }
}
\vskip -20pt
\caption{
 $F_2^c$  as a function of $Q^2$ in GeV 
 computed to ${\cal O}(\alpha_s^1)$ in the 
 ZM-VFN, FFN, ACOT, and SACOT schemes using CTEQ4M PDF's. 
 Fig.~a)  $x=0.1$ , and 
 Fig.~b)  $x=0.001$.
 Figures taken from Ref.~\protect{\cite{sacot}}.
\label{fig:sacot} 
}
\vskip -20pt
\end{figure}
}
\def\figcgn{
\begin{figure}[ht] 
\begin{center}
\leavevmode
 \epsfxsize=0.95\hsize 
\epsfbox{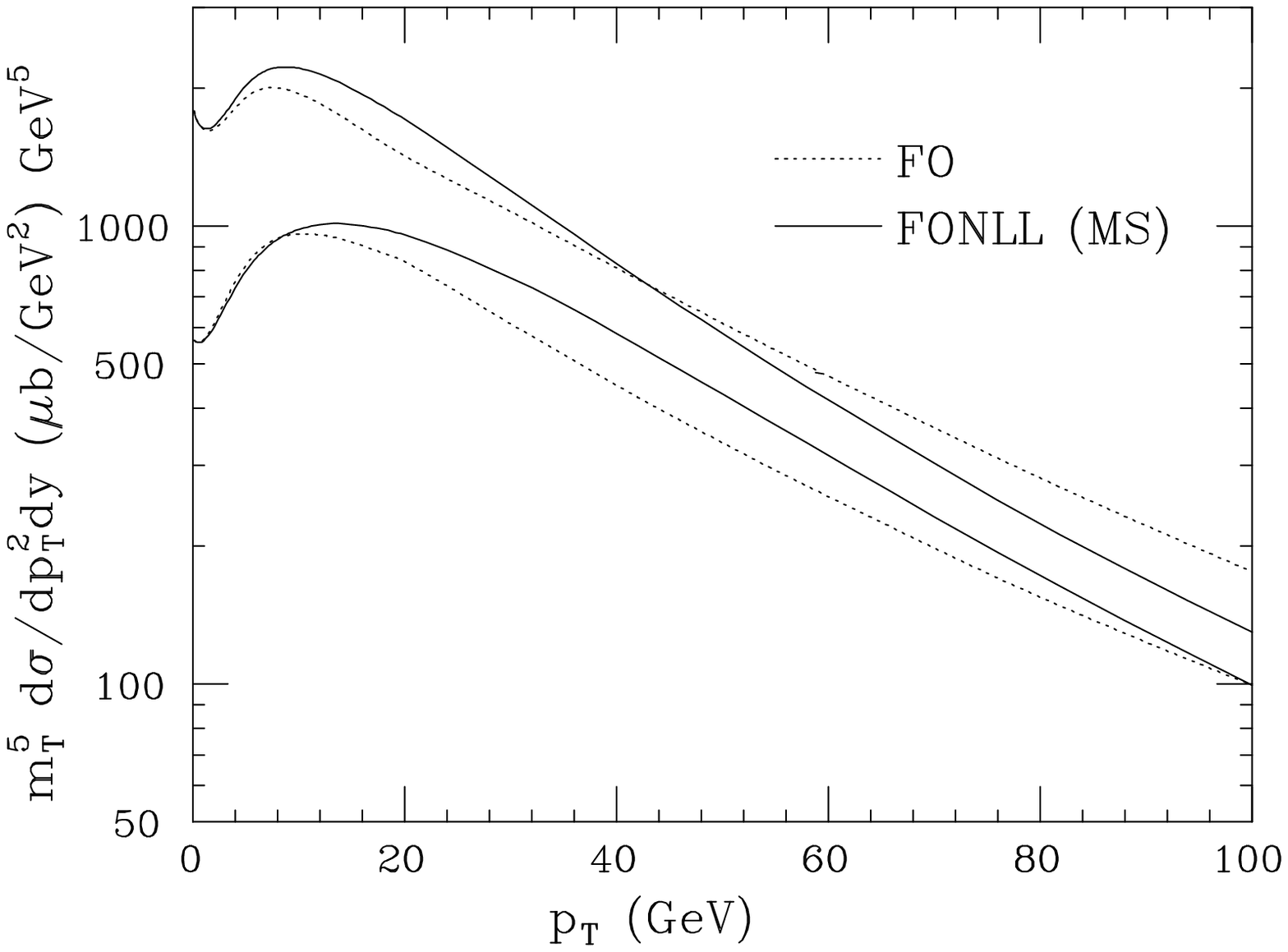} 
\vskip -20pt
\caption{
Differential cross section for b-production {\it vs.} $p_T$
comparing the Fixed-Order (FO) and the Fixed-Order Next-to-Leading-Log
(FONLL) result in the $\overline{MS}$ scheme. 
The bands are obtained by varying independently the renormalization 
and factorization scales. 
 The cross section is scaled by $m_T^5$ with $m_T = \sqrt{m_b^2+p_T^2}$,
and  $\sqrt{s}=1800\, GeV$, $m_b=5\, GeV$, $y=0$, with 
CTEQ3M PDF's. 
 Figure taken from Cacciari,  Greco, and Nason, Ref.~\protect{\cite{cgn}}.  
\label{fig:cgn} 
}
\vskip -20pt
\end{center}
\end{figure}
}
\def\figstk{
\begin{figure}[ht] 
\begin{center}
\leavevmode
 \epsfxsize=0.95\hsize \epsfbox{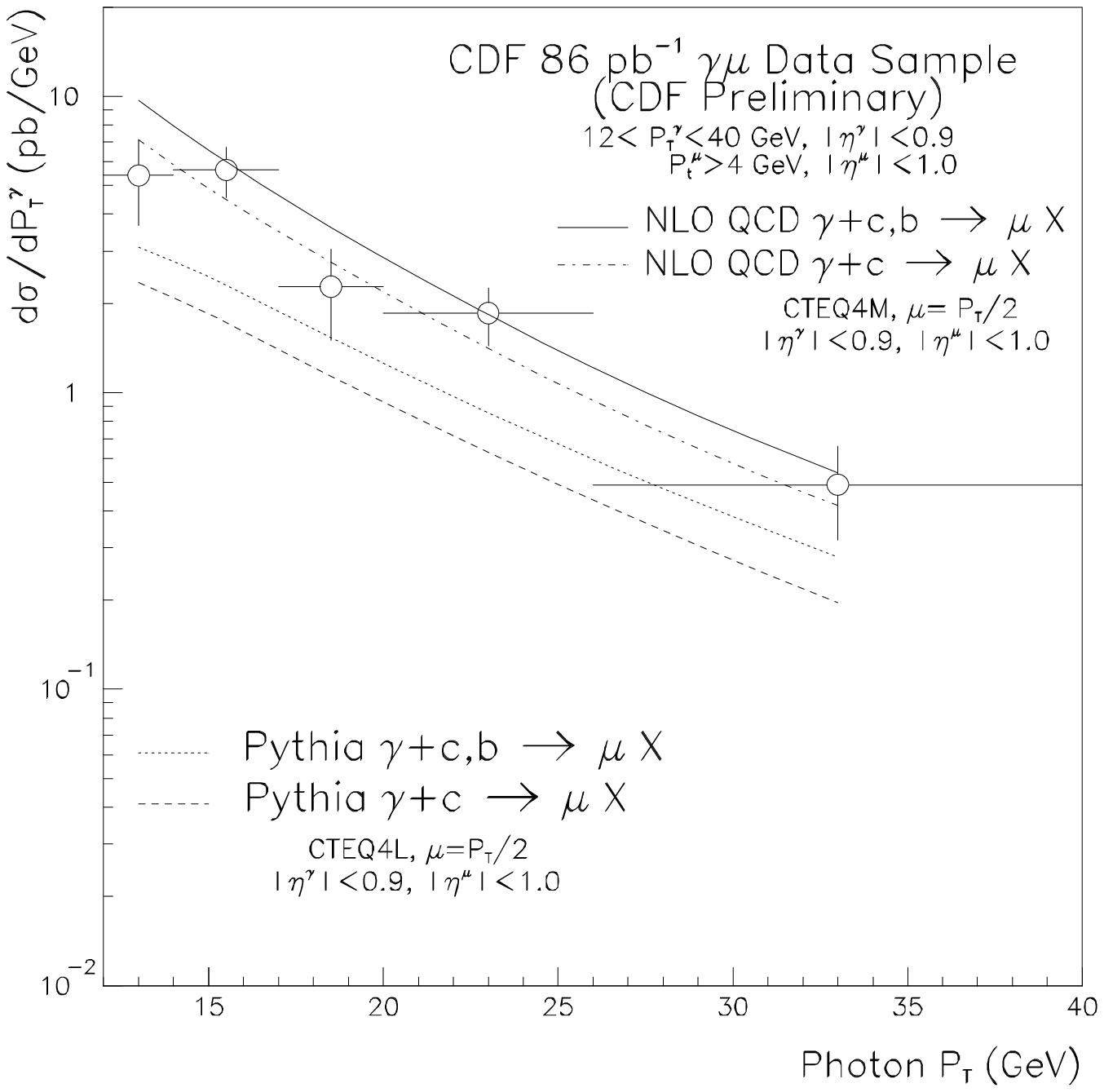} 
\vskip -20pt
\caption{
 Differential $d\sigma/dp_T^{\gamma}$ 
for $\gamma$ plus tagged heavy quark production 
as compared with Pythia and the NLO QCD results. 
 Figure taken from Ref.~\protect{\cite{stk}}.  
\label{fig:stk} 
}
\vskip -20pt
\end{center}
\end{figure}
}
\def\figxsc{
\begin{figure}[ht] 
\begin{center}
\leavevmode
 \epsfxsize=0.95\hsize \epsfbox{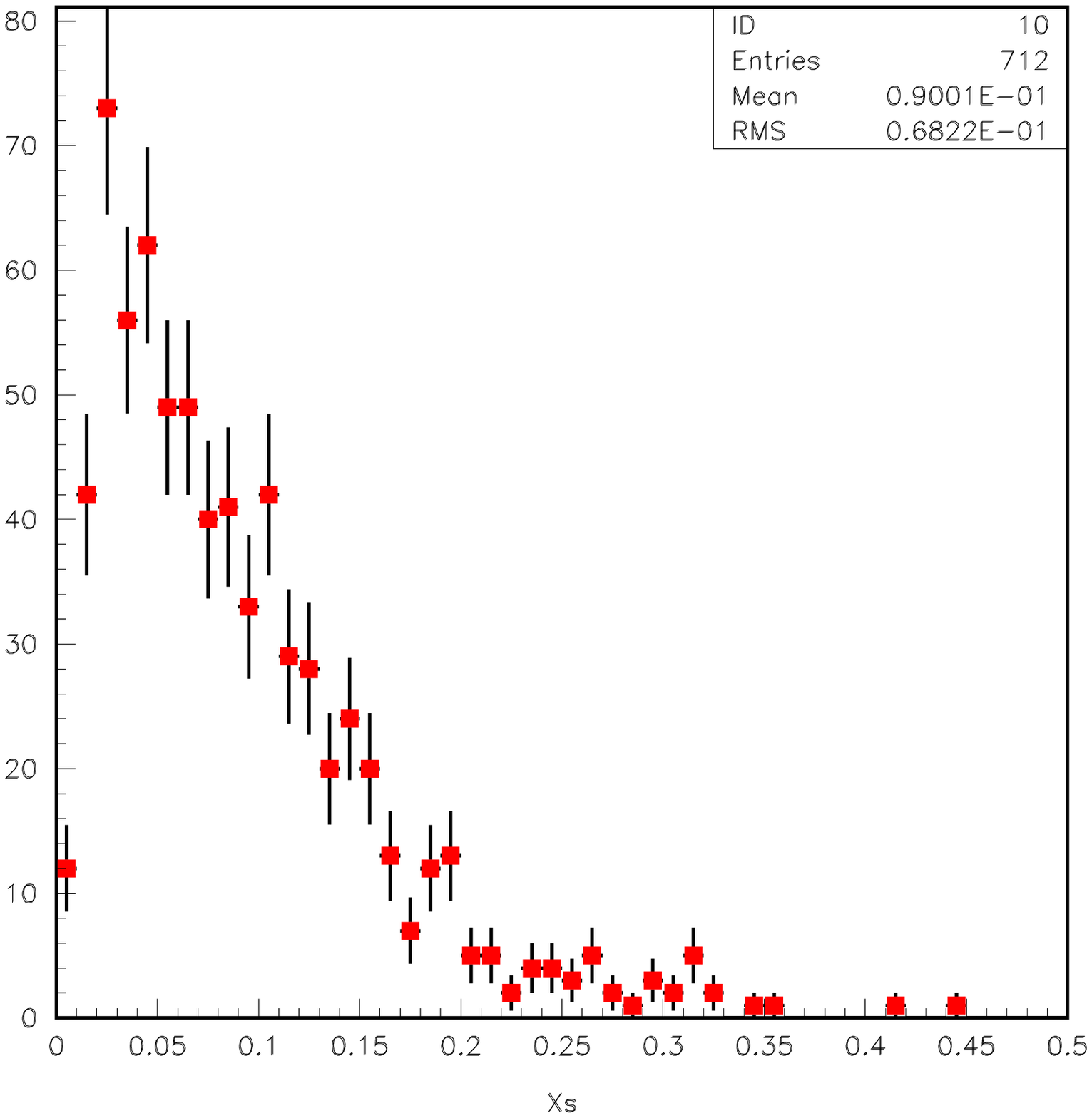} 
\vskip -20pt
\caption{
 Distribution of $Events/0.01$
{\it vs.} $x$ of the s-quarks which contribute to 
the $s+W \to c$ process. 
 Figure taken from Ref.~\protect{\cite{regina}}.  
\label{fig:xsc} 
}
\vskip -20pt
\end{center}
\end{figure}
}
\def\figdff{
\begin{figure}[ht] 
\begin{center}
\leavevmode
\vbox{
 \hbox{
 \epsfxsize=0.85\hsize \epsfbox{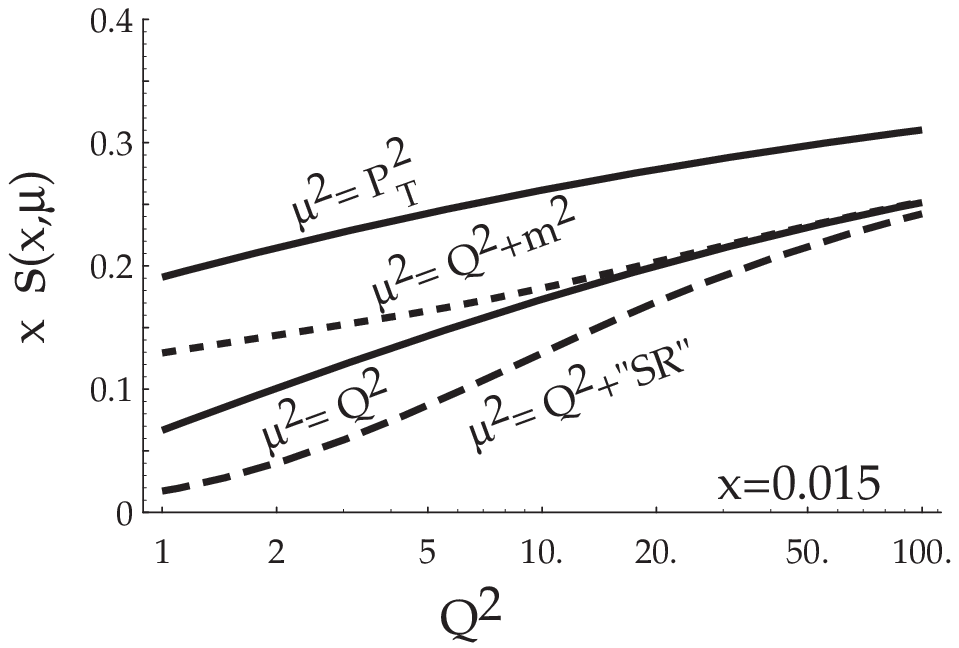} 
 }
 \hbox{
 \epsfxsize=0.85\hsize \epsfbox{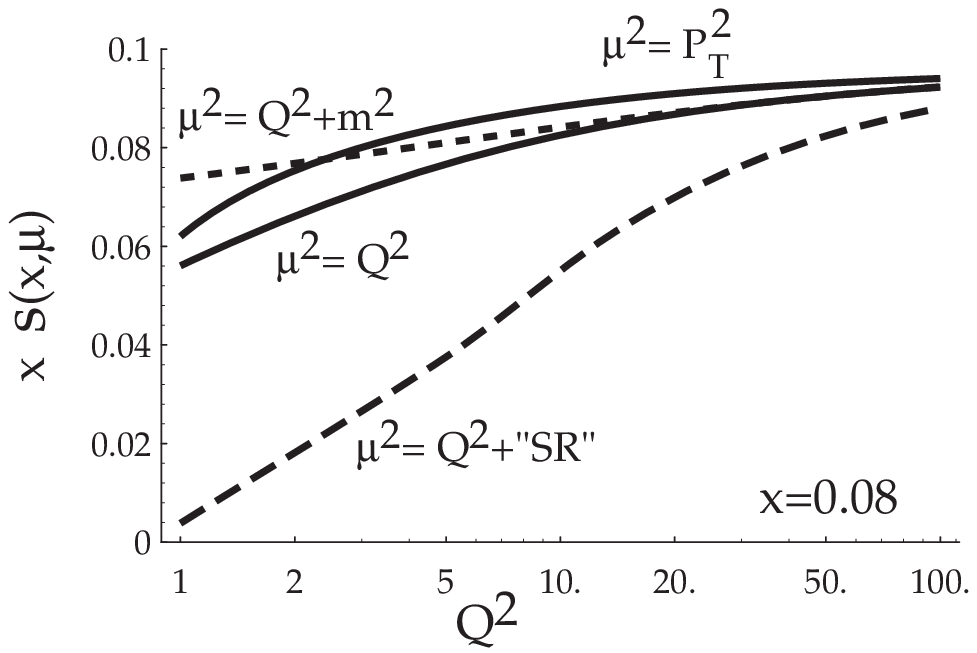}
 }
}
\vskip -20pt
\caption{
 Variation of $x \, s(x,\mu)$ for three choices of $\mu$, and also with
a  ``SR" (slow-rescaling) type correction: $x \to x(1+m_c^2/Q^2)$. 
\label{fig:dff} 
}
\vskip -20pt
\end{center}
\end{figure}
}
\def\figdf{
\begin{figure}[ht] 
\begin{center}
\leavevmode
 \epsfxsize=0.99\hsize \epsfbox{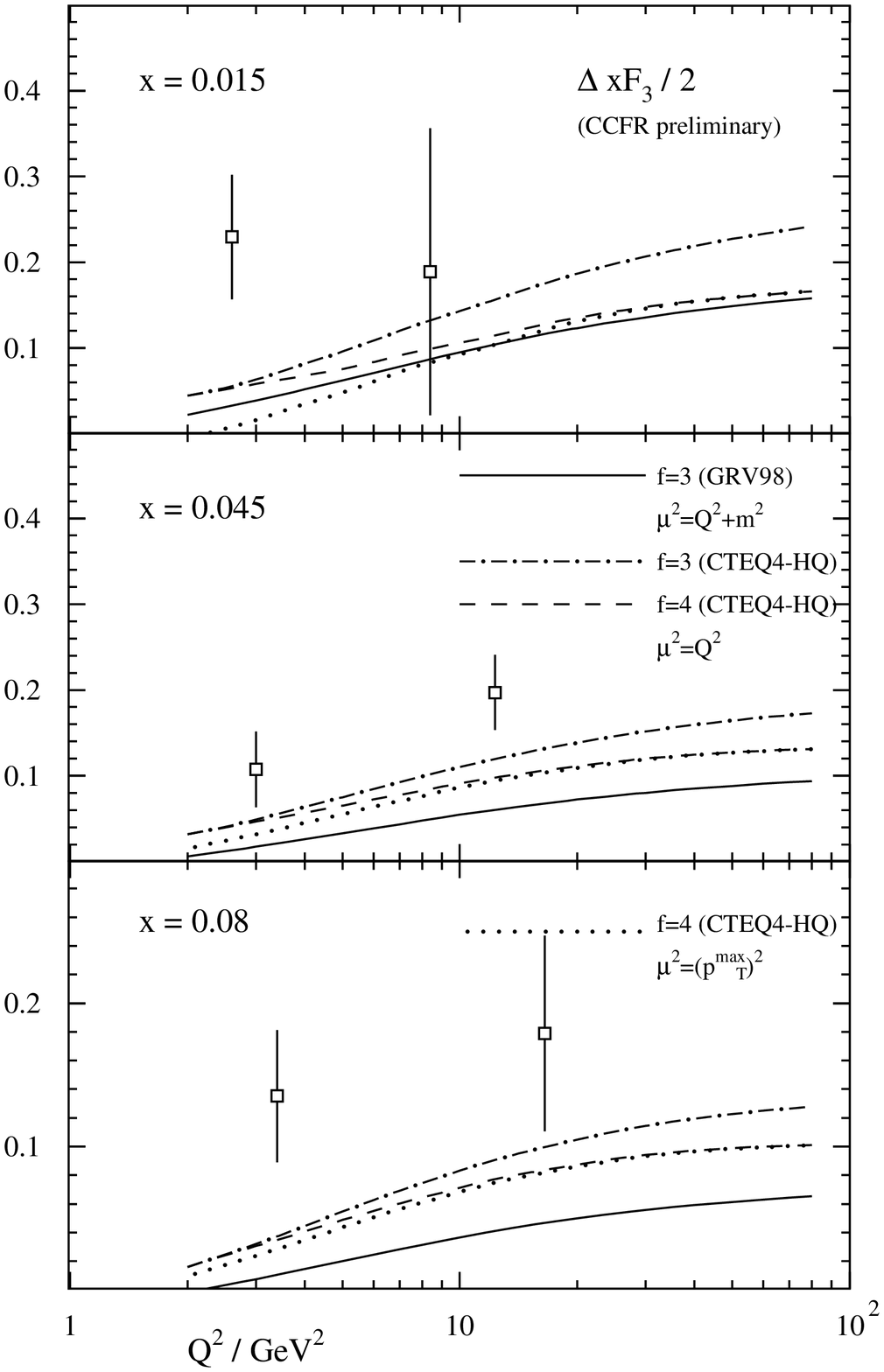} 
\vskip -20pt
\caption{
 $\Delta x F_3/2$ {\it vs.} $Q^2$ for three choices of $x$. 
 Calculations provided by S.~Kretzer. 
\label{fig:df} 
}
\vskip -20pt
\end{center}
\end{figure}
}
\begin{document}
%%%%%%%%%%%%%%%%%%%%%%%%%%%%%%%%%%%%%%%%%%%%%%%%%%%%%%%%%%%%%%%%%%%%%%%%%%%%%%%%

%%%%%%%%%%%%%%%%%%%%%%%%%%%%%%%%%%%%%%%%%%%%%%%%%%%%%%%%%%%%%%%%%%%%%%%%%%%%%%%%
\title{Heavy Quark Production and PDF's Subgroup Report\protect\thanks{%
Contribution to the Physics at Run II Workshops: QCD and Weak Boson Physics,
Fermilab, 1999.
}}
%%%%%%%%%%%%%%%%%%%%%%%%%%%%%%%%%%%%%%%%%%%%%%%%%%%%%%%%%%

\def\addr#1{\address{#1 \vskip -8pt}}

\author{ 
R.~Demina\rlap,\addr{Kansas State University, Physics Department, 116 Cardwell Hall,
  Manhattan, KS 66506}
S.~Keller\rlap,\addr{Theoretical Physics Division, CERN, CH-1211 Geneva 23, 
      	Switzerland}
M.~Kr\"amer\rlap,\addr{Department of Physics, University of Edinburgh, 
       Edinburgh EH9 3JZ, Scotland}
S. Kretzer\rlap,\addr{Univ. Dortmund, Dept. of  Physics, 
      D-44227 Dortmund, Germany}
R.~Martin\rlap,\addr{Univ. of Illinois at Chicago, Dept. of Physics, 
      Chicago, IL 60607}
F.I.~Olness\rlap,\addr{Southern Methodist University, Department of Physics, 
      Dallas, TX 75275-0175}\thanks{Sub-group convenor.}
R.J.~Scalise\rlap,$^{\rm f}$
D.E.~Soper\rlap,\addr{Institute of Theoretical Science, University of Oregon, 
      	Eugene OR 97403, USA}
W.-K.~Tung\rlap,\addr{Michigan State University, Department of Physics and Astronomy, 
      East Lansing, Michigan 48824-1116}%
${}^{,}$\addr{Fermi National Accelerator Laboratory,  
  Batavia, IL 60510}
N.~Varelas\rlap,$^{\rm e}$
U.K. Yang\addr{University of Chicago, Enrico Fermi Institute, 
      Chicago, IL 60637-1434}
}

%%%%%%%%%%%%%%%%%%%%%%%%%%%%%%%%%%%%%%%%%%%%%%%%%%%

\begin{abstract}
We present a status report of a variety of projects related to 
heavy quark production and parton distributions for the 
Tevatron Run~II. 

\end{abstract}

\maketitle

\def\subsection#1{\section{#1}\null \vskip -15pt}
%%%%%%%%%%%%%%%%%%%%%%%%%%%%%%%%%%%%%%%%%%%%%%%%%%%%%%%%%%%%%%%%%%%%%%%%%%%%%%%%
\subsection{Introduction} 
%%%%%%%%%%%%%%%%%%%%%%%%%%%%%%%%%%%%%%%%%%%%%%%%%%%%%%%%%%%%%%%%%%%%%%%%%%%%%%%%

\addtocounter{footnote}{-1}

\addtocounter{footnote}{-1}

The production of heavy quarks, both hadroproduction and leptoproduction, 
 has become an important theoretical and phenomenological
issue. 
 While the hadroproduction mode is of direct interest to this
workshop,\cite{tev} we shall find that the simpler 
leptoproduction process
can provide important insights into the fundamental production 
mechanisms.\cite{nutev,charm,hera,kramer}
 Therefore, in preparation for the Tevatron Run~II, we must consider 
information from a variety of sources including charm and bottom 
production at fixed-target and collider lepton and hadron facilities. 

For example, the charm contribution to the total structure function $F_2$
at
 HERA, is sizeable, up to $\sim 25\%$ in the small $x$ region.\cite{hera}
 Therefore a proper description of charm-quark
production is required for a global analysis of structure
function data,  and hence a precise extraction of the parton densities in
the
proton.
 These elements are important for addressing a variety of issues 
at the Tevatron.

In addition to the studies investigated at the Run~II workshop 
series,\footnote{%
In particular,  in the Run~II B-Physics workshop, 
the studies of {\it Working Group 4: Production, Fragmentation, Spectroscopy}, 
organized by  
Eric Braaten, Keith Ellis, Eric Laenen, William Trischuk, Rick Van Kooten, and Scott Menary,
addressed many issues of direct interest to this subgroup. 
The report is in progress, and the web page is located at: 
\hbox{http://www-theory.fnal.gov/people/ligeti/Brun2/}
}
we want to call attention to the extensive work done in the 
 {\it Standard Model Physics
(and more) at the LHC Workshop} organized by Guido Altarelli, Daniel
Denegri, Daniel Froidevaux, Michelangelo Mangano, Tatsuya Nakada which
was held at CERN during the same period.\footnote{%
The main web page is located at: 
 http://home.cern.ch/$\sim$mlm/lhc99/lhcworkshop.html}
 In particular, the investigations of the LHC {\it b-production group} 
(convenors: Paolo Nason, Giovanni Ridolfi, Olivier Schneider, Giuseppe
Tartarelli, Vikas Pratibha) and the {\it QCD group} (convenors: Stefano
Catani, Davison Soper, W. James Stirling, Stefan Tapprogge, Michael
Dittmar) are directly relevant to the material discussed here.
  Furthermore, our report limits its scope to the issues discussed 
within the Run~II workshop; for a recent comprehensive review,
see Ref.~\cite{Frixione:1997ma}.

%%%%%%%%%%%%%%%%%%%%%%%%%%%%%%%%%%%%%%%%%%%%%%%%%%%%%%%%%%%%%%%%%%%%%%%%%%%%%%%%
\subsection{Schemes for Heavy Quark Production\label{sec:schemes}} 
%%%%%%%%%%%%%%%%%%%%%%%%%%%%%%%%%%%%%%%%%%%%%%%%%%%%%%%%%%%%%%%%%%%%%%%%%%%%%%%%

Heavy quark production also provides an important theoretical challenge as 
the presence of the heavy quark mass, $M$, 
introduces a new scale into the problem. 
 The heavy quark mass scale, $M$, 
in addition to the 
characteristic energy scale of the process 
(which we will label here generically as $E$),
will require a different organization of the perturbation series 
depending on the relative magnitudes of $M$ and $E$. 
  We find there are essentially two cases to 
consider.\footnote{%
We emphasize that the choice of a prescription for dealing with
quark masses in the hard scattering coefficients for deeply inelastic
scattering is a separate issue from the choice of definition of the
parton distribution functions. For all of the prescriptions discussed
here, one uses the standard \MSbar\ definition of parton distributions.}

\begin{enumerate}

\item 
 For the case of $E \sim  M$,  heavy-quark production  is
calculated in the so-called fixed flavor number  (FFN) scheme
from hard processes initiated by light quarks ($u,d,s$) and gluons,
where all effects of the charm quark are contained in the perturbative
coefficient functions. The FFN scheme incorporates the correct
threshold behavior, but for large scales, $E\gg M$, the coefficient
functions in the FFN scheme at higher orders in $\alpha_s$ contain
potentially large logarithms $\ln^n(E^2/M^2)$, which may need to be
resummed.\cite{LRSN,buza,csn,or} 

\item 
 For the case of $E \gg  M$, it is necessary to include the heavy
quark as an active parton in the proton. 
 This serves to resum the potentially large logarithms $\ln^n(E^2/M^2)$
discussed above. 
 The simplest approach
incorporating this idea is the so-called zero mass variable flavor
number  (ZM-VFN) scheme, where heavy quarks are treated as
infinitely massive below some scale $E\sim M$ and massless above this
threshold. This prescription has been used in global fits for many
years, but it has an error of ${\cal O}(M^2/E^2)$ and is not suited
for quantitative analyses unless $E\gg M$.

\end{enumerate}  

While the extreme limits $E \gg  M$ and $E \sim  M$
are straightforward, much of the experimental data lie in the 
intermediate region 
 As such, the correct PQCD formulation of heavy quark production, 
capable of spanning the full energy range,  must incorporate
the physics of both the FFN scheme and the ZM-VFN scheme. 
 Considerable effort has
been made to devise a prescription for heavy-flavor production that
interpolates between the FFN scheme close to threshold and the ZM-VFN
scheme at large $E$. 

The generalized VFN scheme includes the heavy quark as an
active parton flavor and involves matching between the FFN scheme with
three
active flavors and a four-flavor prescription with non-zero
heavy-quark mass. It employs the fact that the mass singularities
associated with the heavy-quark mass can be resummed into the parton
distributions without taking the limit $M\to 0$ in the short-distance
coefficient functions, as done in the ZM-VFN scheme. 
 This is precisely the underlying idea of the
Aivazis--Collins--Olness--Tung
(ACOT) ACOT scheme\cite{acot}
which is based on the renormalization method of 
Collins--Wilczek--Zee (CWZ).\cite{cwz}
 The order-by-order procedure to implement this approach has now
been systematically established to all orders in 
PQCD by Collins.\cite{collins98}

Recently, additional implementations of 
VFN schemes have been
defined in the literature. 
 While these schemes all agree in principle on the
result summed to all orders of perturbation theory,
the way of
ordering the perturbative expansion is not unique and the results
differ at finite order in perturbation theory.
 The Thorne--Roberts (TR) \cite{TR} prescription
has been used in the MRST recent global analyses of parton
distributions.\cite{MRST} 
 The BMSN and CNS prescriptions have made use of the 
${\cal O}(\alpha_s^2)$ calculations by Smith, van~Neerven, and 
collaborators\cite{buza,csn}
to carry these ideas to higher order. 
 The boundary conditions on the PDF's at the
flavor threshold become more complicated at this order; 
in particular, the PDF's are no longer continuous  across the 
N to N+1 flavor threshold. 
Buza {\it et al.},\cite{buza} have computed the matching conditions,
and this has been implemented in an evolution program by CSN.\cite{csn}
 More recently, a Simplified-ACOT (SACOT) 
scheme inspired by the prescription advocated by
Collins~\cite{collins98} was introduced;\cite{sacot}
we describe this new scheme in Sec.~\ref{sec:acot}.

%%%%%%%%%%%%%%%%%%%%%%%%%%%%%%%%%%%%%%%%%%%%%%%%%%%%%%%%%%%%%%%%%%%%%%%%%%%%%%%%
\subsection{From Low To High Energy  Scale}
%%%%%%%%%%%%%%%%%%%%%%%%%%%%%%%%%%%%%%%%%%%%%%%%%%%%%%%%%%%%%%%%%%%%%%%%%%%%%%%%
\figband

To compare the features of the FFN scheme with the ACOT VFN scheme\footnote{%
In this section we shall use the ACOT VFN scheme for this illustration. 
The conclusions extracted in comparison to the FFN scheme are largely independent
of which VFN scheme are used.}
 concretely,
we will take  the example of heavy quark production in DIS; the features 
we extract from this example are directly applicable to the
hadroproduction case
relevant for the Tevatron Run~II.
 One measure we have of estimating the uncertainty of a calculated
quantity
is to examine the variation of the renormalization and factorization 
scale dependence. While this method can only provide a lower bound on the
uncertainty, it is a useful tool.  

In \fig{band}, we display the component of $F_2^{c}$ for the $s+W\to c$
sub-process 
at $x=0.01$ plotted {\it vs.} $Q^2$. We gauge the scale uncertainty by
varying 
$\mu$ from $1/2\, \mu_0$to $2.0\, \mu_0$ with $\mu_0= \sqrt{Q^2+m_c^2}$. 
In this figure, both schemes are applied to ${\cal O}(\alpha_s^1)$.
 We observe  that the FFN scheme is narrower at low $Q$, and increases
slightly
at larger $Q$. This behavior is reasonable given that we expect this 
scheme to work best in the threshold region, but to decrease in accuracy
as the unresummed logs of $\ln^n(Q^2/m_c^2)$ increase. 
 
Conversely, the ACOT VFN scheme has quite the opposite behavior. 
At low $Q$, this calculation displays mild scale uncertainty,
but at large $Q$ this uncertainty is significantly reduced. 
This is an indication that the resummation of the  $\ln^n(Q^2/m_c^2)$ 
terms via the heavy quark PDF serves to decrease the scale uncertainty
at a given order of perturbation theory. 
 While these general results were to be expected, what is surprising
is the magnitude of the scale variation. Even in the threshold region
where
$Q\sim m_c$ we find that the VFN scheme is comparable or better than 
the FFN scheme. 

 At present, the FFN scheme has been calculated to one further order
in perturbation theory, ${\cal O}(\alpha_s^2)$. While the higher order
terms
do serve to reduce the scale uncertainty, it is only at the lowest 
values of $Q$ that the ${\cal O}(\alpha_s^2)$ FFN band is smaller than the
${\cal O}(\alpha_s^1)$ VFN band.
 Recently, ${\cal O}(\alpha_s^2)$ calculations in the VFN scheme 
have been performed;\cite{csn} it would be interesting to extend such
comparisons
to these new calculations.  

Let us also take this opportunity to clarify a misconception that
has occasionally appeared in the literature. The VFN scheme is {\it not} 
required to reduce to the FFN scheme at $Q=m_c$. 
While it is true that the VFN scheme does have the FFN scheme as a limit, 
this matching depends on the definitions of the PDF's, and the 
choice of the $\mu$ scale.\footnote{%
The general renormalization scheme is laid out in the CWZ paper\cite{cwz}.
The matching of the PDF's at ${\cal O}(\alpha_s^1)$ was computed in 
Ref.~\cite{collinstung} and Ref.~\cite{qian}.
The  ${\cal O}(\alpha_s^2)$ boundary conditions were computed in 
Ref.~\cite{buza}.
}
In this particular example, even at $Q=m_c$, 
the resummed logs 
in the heavy quark PDF can yield a non-zero contribution which help
to stabilize the scale dependence of the VFN scheme result.\footnote{%
{\it Cf.}, Ref.~\cite{schmidt} for a detailed discussion.}

The upshot is that even in the threshold region, the resummation of the
logarithms via the heavy quark PDF's can help the stability of the theory.

%%%%%%%%%%%%%%%%%%%%%%%%%%%%%%%%%%%%%%%%%%%%%%%%%%%%%%%%%%%%%%%%%%%%%%%%%%%%%%%%
\subsection{Simplified ACOT (SACOT) prescription\label{sec:acot}}
%%%%%%%%%%%%%%%%%%%%%%%%%%%%%%%%%%%%%%%%%%%%%%%%%%%%%%%%%%%%%%%%%%%%%%%%%%%%%%%%
 \figsacot

We  investigate a modification of the ACOT
scheme inspired by the prescription advocated by
Collins.\cite{collins98} This prescription has the advantage of being
easy to state, and allowing relatively simple calculations. Such
simplicity could be crucial for going beyond one loop order in
calculations.\footnote{%
See Ref.~\cite{sacot} for a detailed definition, discussion, and
comparisons.}

\begin{quote}
{\it Simplified ACOT (SACOT) prescription}.
Set $M_H$ to zero in the calculation of the hard
scattering partonic functions $\widehat \sigma$ for incoming heavy quarks.
\end{quote}

For example, this scheme tremendously simplifies the calculation of 
the neutral current structure function 
$F_2^{charm}$ even at ${\cal O}(\alpha_s^1)$. 
 In  other prescriptions, 
the tree process $\gamma +c \to c+g$ 
and the one loop process $\gamma +c \to c$ must 
be computed with non-zero charm mass, and this results
in a complicated expression.\cite{kretzer}
In the SACOT scheme, the charm mass can be set to zero
so that the final result for these sub-processes reduces 
to the very simple massless result.

While the SACOT scheme allows us to simplify the calculation,
the obvious question is: does this simplified version contain the
full dynamics of the process. 
 To answer this quantitatively,  
we compare prediction for $F_2^{charm}$ obtained with
 1) the SACOT scheme at order $\alpha_s^1$ with 
 2) the predictions obtained with
the original ACOT scheme, 
 3) the ZM-VFN procedure in which the charm quark
can appear as a parton but has zero mass, and 
 4) the FFN procedure in which
the charm quark has its proper mass but does not appear as a parton.
For simplicity, we take $\mu=Q$.

In \fig{sacot} we show $F_2^c(x,Q)$ as a function of $Q$ for $x = 0.1$
and $x = 0.001$ using the CTEQ4M parton distributions.\cite{cteq,LaiTun97a}
 We observe that the ACOT and SACOT schemes are effectively identical
throughout the kinematic range. There is a slight  difference observed in
the threshold region, but this is small in comparison to 
the  renormalization/factorization 
$\mu$-variation (not shown). Hence the difference between the
ACOT and SACOT results is of no physical consequence. The fact that  the
ACOT and SACOT match extremely  well throughout the full kinematic range
provides  explicit numerical verification that the SACOT scheme fully
contains the physics.

Although we have used the example of heavy quark leptoproduction, let
us comment briefly on the implications of this scheme for 
the more complex case of hadroproduction.\cite{tev,nde,Beenakker,ost} 
 At present, we have calculations for the all the ${\cal O}(\alpha_s^2)$ 
hadroproduction sub-processes such as $gg \to Q \bar{Q}$ and $gQ \to gQ$. 
 At ${\cal O}(\alpha_s^3)$ we have the result for the  $gg \to g Q
\bar{Q}$
sub-processes, but not the general result for  $gQ \to g g Q$ with
non-zero heavy quark mass. With the SACOT scheme, 
we can set the heavy quark mass to zero in the  $gQ \to g g Q$ sub-process
and thus make use of the simple result already in the literature.\footnote{%
For a related idea, see the fragmentation function formalism 
of Cacciari and Greco\cite{Cacciari} in the following section.}
 This is just one example of how the SACOT has the practical advantage of 
allowing us to extend our calculations to higher orders in the
perturbation theory. 
 We now turn to the case of heavy quark production for hadron colliders.

%%%%%%%%%%%%%%%%%%%%%%%%%%%%%%%%%%%%%%%%%%%%%%%%%%%%%%%%%%%%%%%%%%%%%%%%%%%%%%%%
\subsection{Heavy Quark Hadroproduction}
%%%%%%%%%%%%%%%%%%%%%%%%%%%%%%%%%%%%%%%%%%%%%%%%%%%%%%%%%%%%%%%%%%%%%%%%%%%%%%%%

\figcgn

There has been notable  progress in the area of 
hadroproduction of heavy quarks. 
The original NLO calculations of the 
$g g \to b \bar{b}$ subprocess were performed by 
 Nason, Dawson, and Ellis \cite{nde},
and by 
Beenakker, Kuijf, van Neerven, Meng, Schuler, and Smith\cite{Beenakker}.
Recently, Cacciari and Greco\cite{Cacciari} 
have used a NLO fragmentation 
formalism to resum the heavy quark contributions in the limit of large $p_T$; 
the result is a decreased renormalization/factorization scale variation in the large $p_T$ region. 
The ACOT scheme was applied to the hadroproduction case by Olness, Scalise, and Tung.\cite{ost}
More recently, the NLO fragmentation 
formalism of Cacciari and Greco has been merged with the massive FFN calculation
of  Nason, Dawson, and Ellis
by Cacciari,  Greco, and Nason,\cite{cgn}; 
the result is a calculation which matches the FFN calculation 
at low $p_T$, and  takes advantage of the NLO fragmentation formalism 
in the high $p_T$ region, thus yielding good behavior 
throughout the full $p_T$ range.
 This is displayed in \fig{cgn} where we see that this 
 Fixed-Order Next-to-Leading-Log (FONLL) calculation 
 displays reduced  scale variation in the large $p_T$ region,
 and matches on the the massive NLO calculation in the small $p_T$ region.
Further details can be found in the report of the LHC Workshop
{\it b-production group}.\footnote{
 The LHC Workshop {\it b-production group} is organized by
 Paolo Nason, Giovanni Ridolfi, Olivier Schneider, Giuseppe
Tartarelli, Vikas Pratibha, and the report is currently in preparation. 
 The webpage for the {\it b-production group} is located at
 http://home.cern.ch/n/nason/www/lhc99/
 }

%%%%%%%%%%%%%%%%%%%%%%%%%%%%%%%%%%%%%%%%%%%%%%%%%%%%%%%%%%%%%%%%%%%%%%%%%%%%%%%%
\subsection{$W$ + Heavy Quark Production}
%%%%%%%%%%%%%%%%%%%%%%%%%%%%%%%%%%%%%%%%%%%%%%%%%%%%%%%%%%%%%%%%%%%%%%%%%%%%%%%%
\tabgkl

\figstk

\figxsc

The precise measurement of $W$ plus heavy quark ($W$+$Q$) events provides
an important 
information on a variety of issues. 
 Measurement of $W$+$Q$ allows us to 
test  NLO  QCD theory  at high scales and  investigate  questions about 
 resummation  and heavy quark PDF's. 
 For example, if sufficient statistics are available,  $W$+$charm$ final
states
can be used to extract information about the strange quark distribution. 
 In an analogous manner, the  $W$+$bottom$ final states 
are sensitive to the charm PDF; 
furthermore, $W$+$bottom$ can fake Higgs events, and are  
also an important background for sbottom ($\widetilde{b}$) searches.

The cross sections for $W$ plus tagged heavy quark jet were computed
in Ref.~\cite{gkl}, and are shown in \tab{gkl}.  
 Note that this process has a large $K$-factor, and hence comparison
between
data and theory will provide discerning test of the NLO QCD theory. 
 While the small cross sections of these channels hindered analysis in
Run~I,
the increased luminosity in Run~II can make this a discriminating tool. 
 For example,  Run~I provided minimal statistics on $W$+$Q$, 
but there was data in the analogous 
neutral current channel $\gamma$+$Q$. 
 \fig{stk} displays preliminary Tevatron data  from Run~I 
and the comparison with both the PYTHIA Monte Carlo and the NLO QCD
calculations;
again, note the large $K$-factor.  If similar results are attainable
in the charged current channel at Run~II, this would be revealing. 

Extensive analysis the $W$+$Q$ production channels were performed in 
Working Group~I:  ``QCD tools for heavy flavors and new physics searches,"
and we can make use of these results
to estimate the precision to which the strange quark distribution can be
extracted. 
We display \fig{xsc} (taken from the WGI
report\cite{regina}) which shows the distribution in $x$ of the s-quarks 
which contribute to the $W$+$c$ process.\footnote{%
For a detailed analysis of this work including 
selection criteria, see the report of 
Working Group~I:  
``QCD Tools For Heavy Flavors And New Physics Searches," 
as well as Ref.~\cite{regina}.}
  This figure indicates that there will good statistics
in an $x$-range comparable to that investigated by 
neutrino DIS experiments;\cite{nutev,charm}
hence, comparison with this data should provide an important test of 
the strange quark sea and the underlying mechanisms for computing 
such processes.

%%%%%%%%%%%%%%%%%%%%%%%%%%%%%%%%%%%%%%%%%%%%%%%%%%%%%%%%%%%%%%%%%%%%%%%%%%%%%%%%
\subsection{The Strange Quark Distribution}
%%%%%%%%%%%%%%%%%%%%%%%%%%%%%%%%%%%%%%%%%%%%%%%%%%%%%%%%%%%%%%%%%%%%%%%%%%%%%%%%
 \figdff

 \figdf

A primary uncertainty for $W$+$charm$ production 
discussed above  comes from the strange sea PDF, $s(x)$,
which has been the subject of controversy for sometime now. 
  One possibility is that new analysis of present data will resolve this
situation
prior to Run~II, and provide precise distributions as an input the the
Tevatron data analysis. 
The converse would be that this situation remains unresolved, in which
case new data
from Run~II may help to finally solve this puzzle.

The strange distribution is directly measured by dimuon production in 
neutrino-nucleon 
scattering.\footnote{%
Presently, there are a number of LO analyses, 
and one NLO  analysis.\cite{nutev,charm}}
 The basic sub-process is $\nu N \to \mu^- c X$ with a subsequent charm
decay $c \to \mu^+ X'$. 

The strange distribution can also be extracted indirectly using a
combination 
of charged ($W^\pm$) and neutral ($\gamma$) current data; however, the systematic uncertainties
involved
in this procedure make an accurate determination
difficult.\cite{yang}
The basic idea is to use the relation
 \begin{equation}
\frac{F_2^{NC}}{F_2^{CC}} =
\frac{5}{18}
\left\{ 
1 - \frac{3}{5} \, \frac{(s+\bar{s})-(c+\bar{c}) + ...}{q+\bar{q}}
\right\}  
\end{equation}
to extract the strange distribution. This method is complicated
by a number of issues including the $xF_3$ component which can 
play a crucial role in the small-$x$ region---precisely
the region where there has been a long-standing discrepancy.

The structure functions are defined in 
terms of the neutrino-nucleon 
cross section via: 
\[
\frac{d^2 \sigma^{\nu,\bar{\nu}}}{dx \, dy}
=
\frac{\scriptstyle G_F^2 M E}{\pi}
\left[ 
F_2 (1-y) + x F_1 y^2 \pm x F_3 y (1-\frac{y}{2}) 
\right]
\nonumber  
\label{eq:}
\]
It is instructive to recall the simple leading-order correspondence 
between the $F$'s and the PDF's:\footnote{%
To exhibit the basic structure, 
the above is taken the limit of 4 quarks, a symmetric sea, and a 
vanishing Cabibbo angle.
 Of course, the actual analysis takes into account the full
structure.\cite{yang}
}
\begin{eqnarray}
F_2^{(\nu,\bar{\nu}) N} &=&
  x 
\left\{ 
u + \bar{u} + d + \bar{d} + 2s + 2c
\right\} 
\nonumber \\
x F_3^{(\nu,\bar{\nu}) N} &=&
 x 
\left\{ 
u - \bar{u} + d - \bar{d} \pm 2s \mp 2c
\right\} 
\label{eq:}
\end{eqnarray}
 Therefore, the combination $\Delta xF_3$:
\begin{equation}
\Delta xF_3 = x F_3^{\nu N} - x F_3^{\bar{\nu} N} =
4 x \{ s - c \}
\label{eq:}
\end{equation}
 can be used to
probe the strange sea distribution, 
and to understand heavy quark (charm) production. 
 This information, together with the exclusive dimuon events,
may provide a more precise determination of the strange quark sea. 

To gauge the dependence of $\Delta xF_3$ upon various factors, we first 
consider $x s(x,\mu)$  in \fig{dff}, and then the full NLO $\Delta xF_3$
in \fig{df}; this allows us to see the connection between $\Delta xF_3$ and
$x s(x,\mu)$ beyond leading order. 
 In \fig{dff} we have plotted the quantity $x s(x,\mu)$  
{\it vs.} $Q^2$ for two choices of $x$ in a range relevant to the
the dimuon measurements. 
 We use three choices of the $\mu^2$ scale: $\{Q^2, Q^2+m_c^2, P_{T_{max}}^2  \}$.
The choices $Q^2$ and $Q^2+m_c^2$ differ only at lower values of $Q^2$; 
the choice $P_{T_{max}}^2$ is comparable to $Q^2$ and $Q^2+m_c^2$ at $x=0.08$
but lies above for $x=0.015$. 
The fourth curve labeled $Q^2+``SR"$ uses $\mu^2=Q^2$ with a ``slow-rescaling" 
type of correction which (crudely) includes mass effects by 
shifting $x$ to $x(1+m_c^2/Q^2)$; note, the result of this 
correction is significant at large $x$ and low $Q^2$. 

In \fig{df} we have plotted the quantity $\Delta xF_3/2$ 
for an isoscalar target computed to 
order $\alpha_s^1$.
 We display three calculations for three different $x$-bins relevant to 
strange sea measurement.
 1)~A 3-flavor calculation using the GRV98\cite{grv98} distributions,\footnote{%
 The scale choice $\mu=\sqrt{Q^2+m^2}$ for the 3-flavor GRV calculation 
precisely cancels the
collinear strange quark mass logarithm in the coefficient function
thereby making the coefficient function an exact scaling 
function, {\it i.e.} independent of $\mu^2$.}
 and $\mu=\sqrt{Q^2+m^2}$. 
 2)~A 3-flavor calculation using the CTEQ4HQ distributions, and $\mu=Q$. 
 3)~A 4-flavor calculation using the CTEQ4HQ distributions, and $\mu=Q$. 

The two CTEQ curves show the effect of the charm distribution, and
the GRV curve shows the effect of using a different PDF set.
Recall that the GRV calculation corresponds to a FFN scheme.

The pair of curves using the CTEQ4HQ distributions nicely illustrates
how the charm distribution $c(x,\mu^2)$ evolves as $\ln(Q^2/m_c^2)$ for
increasing $Q^2$; note, $c(x,\mu^2)$ enters with a negative sign 
so that the 4-flavor result is below the 3-flavor curve. 
 The choice $\mu=Q$ ensures the 3- and 4-flavor calculation coincide
at $\mu=Q=m_c$; while this choice is useful for instructive purposes, 
a more practical choice might be $\mu \sim \sqrt{Q^2+m^2}$, 
{\it cf.}, Sec.~\ref{sec:schemes}, and Ref.~\cite{schmidt}.
 
For comparison, we also display preliminary data from the 
CCFR analysis.\cite{yang} While there is much freedom in the
theoretical calculation, the difference between these calculations
and the data at low $Q$ values warrants further investigation.

%%%%%%%%%%%%%%%%%%%%%%%%%%%%%%%%%%%%%%%%%%%%%%%%%%%%%%%%%%%%%%%%%%%%%%%%%%%%%%%%
\subsection{Conclusions and Outlook}
%%%%%%%%%%%%%%%%%%%%%%%%%%%%%%%%%%%%%%%%%%%%%%%%%%%%%%%%%%%%%%%%%%%%%%%%%%%%%%%%

A detailed understanding of heavy quark production and heavy quark PDF's
at the Tevatron Run~II will require analysis of fixed-target and HERA
data as well as Run~I results.  
 Comprehensive analysis of the combined data set can provide incisive tests
of the theoretical methods in an unexplored regime, and enable precise
predictions that will facilitate new particle searches in a variety of channels. 
 This document serves as a progress report, and work on these 
topics will continue in preparation for   the Tevatron Run~II.

%%%%%%%%%%%%%%%%%%%%%%%%%%%%%%%%%%%%%%%%%%%%%%%%%%%%%%%%%%%%%%%%%%%%%%%%%%%%%%%%
%\section{Acknowledgments} 
%%%%%%%%%%%%%%%%%%%%%%%%%%%%%%%%%%%%%%%%%%%%%%%%%%%%%%%%%%%%%%%%%%%%%%%%%%%%%%%%

 This work is supported
by the U.S. Department of Energy, 
the National Science Foundation, 
and the Lightner-Sams Foundation.

%%%%%%%%%%%%%%%%%%%%%%%%%%%%%%%%%%%%%%%%%%%%%%%%%%%%%%%%%%%%%%%%%%%%%%%%%%%%%%%%

%%%%%%%%%%%%%%%%%%%%%%%%%%%%%%%%%%%%%%%%%%%%%%%%%%%%%%%%%%%%%%%%%%%%%%%%%%%%%%%%

%%%%%%%%%%%%%%%%%%%%%%%%%%%%%%%%%%%%%%%%%%%%%%%%%%%%%%%%%%%%%%%%%%%%%%%%%%%%%%%%
\end{document}